\documentclass[journal,transmag]{IEEEtran}
    \usepackage{graphicx}
    \usepackage{subcaption}
    \usepackage{caption}
\ifCLASSINFOpdf
\else
\fi
\begin{document}
%
\title{Malware Detection and Analysis: Challenges and Research Opportunities}
%
%
\author{\IEEEauthorblockN{Zahid Akhtar}\vspace{1mm}
\IEEEauthorblockA{Department of Network and Computer Security,\\ State University of New York Polytechnic Institute, USA.\\ \small Email: akhtarz@sunypoly.edu}}

\maketitle

\begin{abstract}
Malwares are continuously growing in sophistication and numbers. Over the last decade, remarkable progress has been achieved in anti-malware mechanisms. However, several pressing issues (e.g., unknown malware samples detection) still need to be addressed adequately. This article first presents a concise overview of malware along with anti-malware and then summarizes various research challenges. This is a theoretical and perspective article that is hoped to complement earlier articles and works.
\end{abstract}


%
\IEEEpeerreviewmaketitle

\section{Introduction}
\label{sec:introduction}

Use of personal computers and mobile devices coupled with internet has now become integral part of everyday life. This ubiquity of high interconnectivity has prompted many serious privacy and security menaces as well as different other malicious activities. For instance, 117 million LinkedIn user's email and password were made publicly available by hackers in 2016. In 2017, Uber revealed that its server was attacked and 57 million drivers and riders data were stolen. While, in 2018 almost 50 million Facebook accounts were compromised due to security breach. Similarly, cyberattacks on Norway’s `Health South East RHF' healthcare organization in 2018 exposed health record of more than half of country's population. Moreover, it is estimated that on an average every 10 second a new malicious code specimen is released to attack mobile devices \cite{Faruki2015}. A surge of cyberattacks with increasing number and sophistication can be seen with each passing year, which is impacting governments, enterprises and individual alike and causing severe reputation, financial and social damages. For example, malicious cyber activities in 2016 cost U.S. economy alone up to 109 billion USD \cite{Cost2018}.

Different types of cyberattacks are presently being performed by cybercriminals, e.g., man-in-the-middle, malware or birthday attack. In particular, malware attacks have advanced as one of the main formidable issues in cybersecurity domain as well as primary tool utilized by cybercriminals \cite{Souri2018}. Malware is a short form of \emph{mal}icious \emph{soft}ware. In French language, `mal' means `bad'. Malware is a catch-all term widely employed to denote various kinds of unwanted harmful software programs with malicious motives \cite{Ucci2017}. When malware is executed on a system or computing device it attempts to breach the system/device's security policies regarding integrity, confidentiality and availability of data. Other names for malware are badware, malicious code, malicious executable and malicious program. Malwares are developed or used by hobbyists and cyber-offenders trying to show their ability by causing havoc and to steal information potentially for monetary gains, respectively. They are popularly known as hackers, black hats and crackers, and could be external/internal menace, industrial spies or foreign governments. Malwares can be used to change or erase data from victim computers, to collect confidential information, or to hijack systems in order to attack other devices, send spams, host and share illicit contents, bring down servers, penetrate networks, and cripple critical infrastructures.

Consequently, a broad range of tools and schemes have been devised to detect and mitigate malware attacks \cite{Faruki2015}. Anti-malware systems thwart malwares by determining whether given program has malign intent or not \cite{Ucci2017}. Despite great advancement of malware defense techniques and their incessant evolution, badwares still can bypass the anti-malware solutions owing to mainly sophisticated packers and weakest link, i.e., humans. Namely, most anti-malware methods do not exhibit low enough error rates. Additionally, their performance particularly drops when they face unknown malwares. While, daily 360,000 novel malware samples hit the scene \cite{Ucci2017}. As anti-malware becomes more avant-garde so as malwares in the wild, thereby escalating the arms race between malware guardians and writers. The quests for scalable and robust automated malware detection frameworks still have to go a long way. This article presents an overview of malwares and their defenses formulated in recent years, and highlights challenges, open issues and research opportunities for researchers and engineers. It is a perspective and academic article, which is aimed at complementing existing  studies and prompt interdisciplinary research.

\section{Malware Categories}
\label{sec:categories}
\begin{figure*}[t]
\begin{center}
\includegraphics[width=\textwidth]{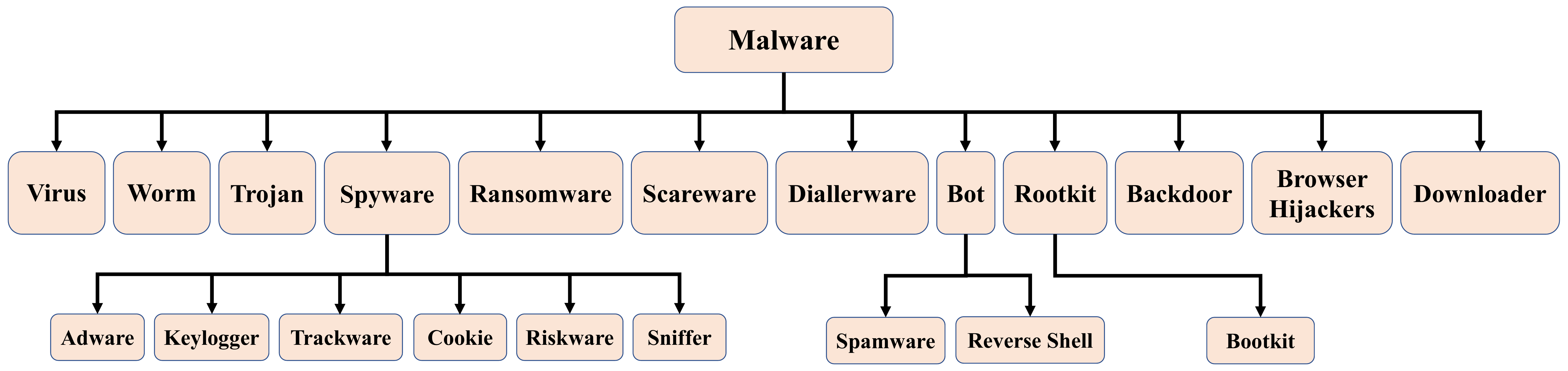}
\caption{General taxonomy of malware.}
\label{fig:taxonomy}
\end{center}
\end{figure*}

Malwares, as depicted in Fig. \ref{fig:taxonomy}, can be divided into various types depending on how they infect, propagate or exploit the target system as follows \cite{Souri2018}. Please note that some of the malware types/tools/names fall in gray area of features intended for begin purposes as well, e.g., cookie, Wireshark, etc.

\subsection{Virus} 
A piece of code that duplicates, reproduces or propagates itself across files, programs and machines if they are network-connected. Viruses cannot execute independently, therefore they are mainly appended to `host' programs/files (e.g., executable files, master boot record). When executed by `host' can corrupt or destroy files, programs, computer's functioning and shared network that may result in denial of service and system's performance degradation. Examples of viruses are Melissa virus and Creeper virus.

\subsection{Worm} 
Unlike virus, worm does not need `host' but can run independently. Worms are self-replicating and self-propagating via a network or storage devices. Worms exploit operation system vulnerabilities, but do not corrupt user or system files. They consume computing and network resources by residing in main memory while replicating and spreading themselves causing DoS and SPD. Examples are MyDoom and SQL Slammer. 

\subsection{Trojan} 
Trojan surfaces as benign program but performs malevolent activities in the backend without user's knowledge. Trojans usually do not infect files or replicate themselves, rather create backdoors for unauthorized system access to delete files, install programs or extricate private data (e.g., passwords). Examples are Zeus and Zitmo.

\subsection{Spyware} 
Spyware spies on users without their knowledge or consent. It is generally used to surveil user activities, gather keystrokes or harvest sensitive information (e.g., login credentials). Examples of spyware are LogKext and GPSSpy. Following are popular spyware sub-categories: 

\subsubsection{Adware}
Adware dispenses either spiteful code/content or ads to infected users via web browser, mobile app or PC’s desktop to generate attacker’s revenue. Another name of this malware is malvertising, as it may use reputed companies/banners to distribute malicious codes. It can be considered as a subcategory of spyware, but unlikely leading to a big harm until coupled with other spywares. Examples are AllSearchApp and Plankton. 

Pornware is also seen as a subclass of adware, when installed maliciously without user’s knowledge to display pornographic materials. 

\subsubsection{Keylogger}
This malware is also called keystroke logger, password grabbers, sniffer or information-stealing malware, which is employed by attackers to record each keystroke to steal sensitive data (e.g., passwords, credit card numbers). Keylogger is generally transferred to a system when spiteful-software is installed or -site is visited. Examples are SpyEye and Formgrabber. 

\subsubsection{Trackware}
Trackware is unwanted software that tracks and collects user activities and habits then share data with a third party. Though trackwares harm user’s privacy, they do not harvest confidential or personally identifiable information. Examples are Trackware. Rewardnet and Win32/WebHancer.A.

\subsubsection{Cookie}
Cookies are plain text files with information of user's web browsing sessions. They are stored on user's computer/device for future use. Although cookies seemingly are not detrimental, they can be menace if exploited by some spyware. Likewise, tracking cookies can be utilized by hackers to gain user's personal details.

\subsubsection{Riskware}
Riskware (aka grayware) is a genuine program when utilized by the attacker can cause damage to system security or data by deletion, duplication, or modification. Authentic programs for riskware could be IRC client, file downloaders, etc.  

\subsubsection{Sniffer}
It is a malicious program that observes and captures the network traffics. It analyzes various fields of packets and gather data for preparation of malware attacks. Sniffers can be `Ethereal' (i.e., legitimate used for troubleshooting) and `BUTTSniffer' (i.e., illegitimate for malign purposes). Examples of sniffers are Wireshark and Aircrack-ng.

\subsection{Ransomware} 
Ransomwares are covertly installed on a victim computer to execute a cryptovirology attack. This malware type encrypts the data or locks down the system, thereby restricting user access till ransom is paid. Specifically, ransomwares can be classified in two main groups, viz. locker ransomwares that decline access to the system/device functionalities, and crypto ransomware that avert access to files/data. Ransomwares examples are FakeDefender and TorrentLocker. 

\subsection{Scareware} 
Scareware deludes people into buying/downloading inessential and potentially perilous security software, opening attachments or visiting a malevolent website. It mostly attempts to frighten users (e.g., by displaying false warning messages), and when installed it collects stored information from victim system, which maybe sold to cybercriminals. Examples are Mac Defender and Registry Cleaner XP. 

\subsection{Diallerware}
Diallerware send premium-rate SMS/multimedia messages without mobile user’s knowledge, thereby causing monetary sums to user. The premium-rate SMS/multimedia messages/calls provide value-added services, which can be abused by attackers. Attackers lure mobile owners to sign up for the premium services managed by themselves, e.g., HippoSMS. Diallerware blocks the messages from service providers to users to avoid user’s awareness of unwanted additional charges.

\subsection{Bot}
A bot (abbreviated from robot) is a malicious program, which enables attacker (aka botmaster or bot herder) to remotely control infected machine without user's knowledge via a command and control (C\&C) channel from a system called C\&C server. A cluster of bots controlled by a sole server is known as botnet. Botnets can be employed to organize DDoS attacks, phishing fraud, sending spams, etc. Well-known examples are Sdbot and Agobot.

\subsubsection{Spamware} Spamware (aka spam sending malware or spambot) is malicious software designed to search and compile list of email addresses as well as sending large number of spam emails. It is an element of a botnet functioning as a distributed spam-sending network. Spamware can use infected user's email ID or IP address to send emails, which may consume great amount of bandwidth and slow down the system. Examples are Trik Spam and Necurs botnet.

\subsubsection{Reverse Shell} 
A reverse shell is an unauthorized program (malware) that provides access of undermined computer to the attacker. Reverse shell enables attacker to run and type command on host as the attacker is local. Examples are Netcat and JSP web shell.

\subsection{Rootkit}
A rootkit is a stealthy software that is devised to conceal specific programs/processes and enabling privileged access to computer/data. Rootkit allows the attacker accessing and controlling the system remotely without being detected, as it normally runs with root privileges and subverts system logs and security software. Examples are NTRootkit and Stuxnet.

\subsubsection{Bootkit}
Bootkit is an advanced form of rootkits that infects master boot record or volume boot record. Since it resides in boot sector, it is difficult to be detected by security software, and also stays active after system reboot. Well-known examples are BOOTRASH and FinFisher.

\subsection{Backdoor}
Backdoor is a malware that installs by itself and creates secret entrance for attackers to bypass system's authentication procedures and to access and perform illegitimate activities. Backdoors are never utilized alone but as foregoing malware attacks of other kinds, as they do not harm but furnish wider attack surfaces. A notable backdoor tool is Remote Access Terminal/Trojan (RAT). Other examples are Basebridge and Olyx.

\subsection{Browser Hijackers}
It is an undesired software that alters settings of web browser without user’s consent either to inject ads in the browser or replace home/error page and search engine. Some of them may access sensitive data with spyware. Examples are CoolWebSearch and RocketTab.

\subsection{Downloader}
It is a malicious program that downloads and installs/runs new versions of malwares from internet on compromised computers. Downloader is usually embedded in websites and software. Examples are Trojan-Downloader:W32/JQCN and Trojan-Downloader:OSX/Jahlev.A.

\begin{figure*}[t]
\begin{center}
\includegraphics[width=0.95\textwidth]{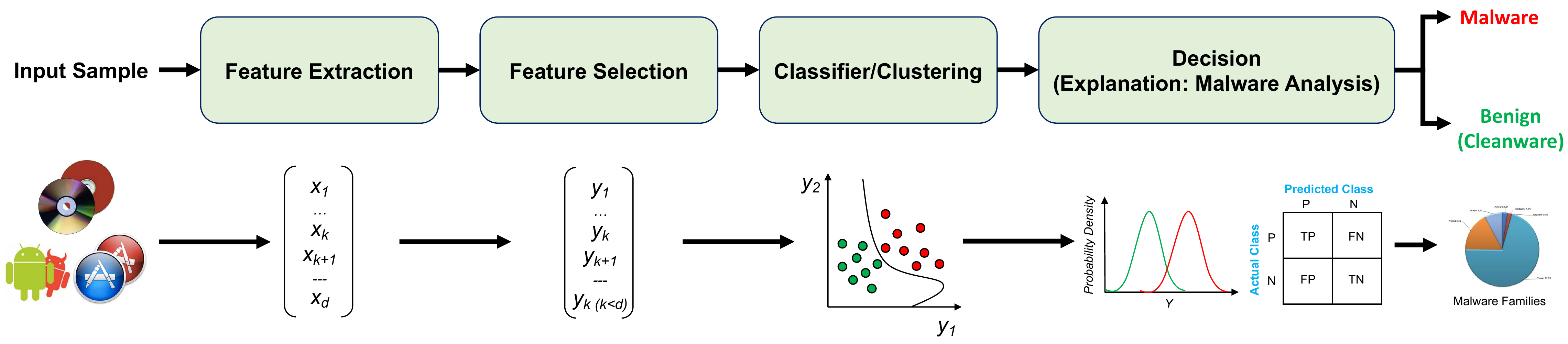}
\caption{A generic malware detection and analysis system. First, input sample is provided to feature extraction module that yields feature representation vector. A feature reduction/selection process is carried out on feature representation vector to obtain fixed dimensionality regardless of length of input sample for enhanced performance. A classification/clustering technique is trained on available set of malware and benign samples. During detection/analysis, unseen sample is reported by the classification/clustering techniques as malware or not. Further analysis is also sometimes performed, e.g., describing suspicious (or benign) characteristics present in the sample.}
\label{fig:detector}
\end{center}
\end{figure*}

\section{Malware Concealment Techniques}
\label{sec:malwareconcealmenttechniques}

To evade anti-malwares, malware writers have applied following different malware camouflage approaches \cite{Souri2018}:

\subsection{Encryption}
Encrypted malware by this method consists of encryption and decryption algorithms, keys and malicious codes. Each time attacker employs new encryption algorithm and key to generate novel malware version. Since decryption algorithm remains same, there is a higher probability to be detected. The main target of this procedure is to avoid static analysis and delaying investigation process. CASCADE was reported as the first encrypted malware in 1987.

\subsection{Packing}
Packing mechanism is utilized to compress/encrypt malware executable file. To detect malwares with packing technique, reverse engineering methods or correct unpacking algorithm is needed, which sometime is hard as it requires knowledge of true packing/compression algorithm. UPX and Upack are examples of packing. 

\subsection{Obfuscation}
This technique obscures program's principal logic to stop others gaining associated knowledge of the code. Malwares with obfuscation and their deleterious functionality stay unintelligible till activated. Quintessential obfuscation strategies are inessential jumps and including garbage commands.

\subsection{Polymorphism}
Polymorphic malware is designed to alter its appearance every time it is executed while retaining original code entirely. Compared to encryption technique, boundless number of encryption algorithms can be utilized by a polymorphic malware such that in each implementation a decryption code's portion is mutated. Transformation engine is generally kept in encrypted malware. When any mutation occurs, a random encryption algorithm is produced to re-encrypt the engine and malware with new decryption key. Different inimical actions can be embedded under encryption operations. Since original code remains intact, polymorphic malwares comparatively become easy to be detected. First known polymorphic virus developed in 1990 is 1260.

\subsection{Metamorphism}
Metamorphism malware (aka body-polymorphic malware) mutates its malevolent codes in each execution to create novel instance that has no similitude with native codes, but functionality yet remains the same. There are two categories of metamorphic malwares. \emph{Open-world malware} that mutates by communicating with other sites over net, e.g., Conficker worm. \emph{Open-world malware} that reprograms itself without external communication by mutating binary code (i.e., binary transformer) or employing pseudocode representation, e.g., Win32/Apparition virus.

\section{Malware Detection and Analysis System}
\label{sec:malwaredetection}

As Fig. \ref{fig:detector} shows, a generic malware detection system consists of four main modules: feature extraction, feature selection, classification/clustering, and decision. The raw data sample is input to feature extraction module, which extricates salient attributes as a feature set. Next, feature selection is performed to tackle the curse of dimensionality, to reduce the computational complexity, and to increase performance of the system by quantifying feature correlations. The resultant feature vector is given to a classifier/clustering scheme. Finally, decision module is employed either to acquire the final binary decision: malware or benign (cleanware), or/and for additional malware analysis such as malware variants detection (i.e., recognizing variant and families), malware category detection (i.e., categorizing based on malwares' prominent behaviors and objectives), malware similarity and novelty detection (i.e., acquiring knowledge about unknown sample by specific similarities and differences against known ones), malware development detection (i.e., finding out if the malware writer has previously submitted it to online defense tools), and malware attribution (i.e., identifying its programming language, from where its launched and actor/group involved).

\subsection{Malware Analysis}
In general, malware analysis is deployed both for detecting/classification and other investigations (e.g., understanding the working to devise novel identification schemes) of malware. Different features such as strings (i.e., frequency of code fragments, names, etc.), byte sequences (i.e., characterization of byte-level contents), opcodes (i.e., identification of machine/assembly-level operations), system/APIs calls (i.e., analyses of execution traces/disassembly code or characterization of APIs' executed actions), call graphs and data dependent (i.e., analyses of data being exchanged between process calls), control flow graphs (i.e., behavior relationships of data flow between system resources), multilayer dependency chains (i.e., characterization of sub-behaviors to capture interactions among samples and system levels), causal dependency graphs (i.e., tracking persistent state changes on target system), influence graphs (i.e., encoding of downloads by malware), memory accesses (i.e., analyses of memory during malware executions), file system (i.e., frequency of created/deleted/modified files), system registry (i.e., count of queried/deleted/modified registry keys), CPU registers (i.e., frequency of registers usages/changes), function length (i.e., number of bytes in a function), exceptions (i.e., exceptions prompted during malware execution) and network traffic (i.e., analyses of incoming and outgoing packets, visited addresses, etc.) are being used for malware analysis. Malware analysis can be conducted in following three ways:

\subsubsection{Static Analysis}

It is also called signature-based, code analysis, white-box or misuse detection approach. Methods in this category generally review statically the code-structure for traits of infections using a pre-defined list of known assails’ signatures without executing the sample \cite{Ucci2017}. However, advanced static analysis techniques may run the sample by deploying reverse engineering, i.e., obtaining binary and assembly codes using decompiler, disassembler and debugger.

Hellal \emph{et al.} \cite{Romdhane2016} presented a call code graph mining based static analysis scheme, called minimal contrast frequent subgraph miner algorithm, to distinguish variants of malware in Windows environment. Schultz \emph{et al.} \cite{Schultz2001} used features like list of DLLs functions, system calls and hex-dump to detect novel and unseen malicious executables. Martin \emph{et al.} \cite{Martin2016} designed a malware detection method that uses third-party API calls in Java files and multi-objective optimization classification. While, Narayanan \emph{et al.} \cite{Narayanan2017} developed a mutli-view (MKLDROID) framework utilizing a graph kernel with multiple kernel learning to determine sets of semantic structures and contextual information from Android apps for malware/malicious code localization. Yerima and Sezer \cite{Yerima2018} proposed Android malware detection that analyzes permissions and intents from the apps via multilevel classifier rank fusion architecture. Recenlty, Cakir \emph{et al.} \cite{Cakir2018} designed a shallow deep learning based method that employed word2vec features via opcodes and a Gradient boosting classifier.

Though static analysis techniques are capable of fast recognizing malwares in versatile applications and pose no risk of infection while analyzing malwares, they need huge number of pre-defined signature dataset. Moreover, they suffer from runtime overhead and cannot discriminate variations of known- or obscure-malwares and zero-day intrusions.

\subsubsection{Dynamic Analysis}

It is also called behavior-based, behavioral analysis, anomaly-based, specification-based or black-box approach. Methods in this category assess samples via their activities by executing them in a confined/simulated environment, e.g., sandboxed, simulator, debugger, virtual machine or emulator.

Miao \emph{et al.} \cite{Miao2016} proposed a bi-layer behavior abstraction technique via semantic examination of dynamic API sequences in Windows environment. Lower- and higher-layer behaviors were captured using data dependence of APIs and complex good interpretability of lower abstractions, respectively. In \cite{Nikolopoulos2016}, authors developed a graph-based model harnessing relations (i.e., dependency graphs) among system-calls' groups for smartphone malicious software detection, but the model requires high time consumption. Authors in \cite{Wuechner2017} presented a compression-based feature mining on system/API calls' quantitative information flow graphs to detect Windows malware. Mao \emph{et al.} \cite{Mao2017} designed a security dependence network from access behaviors to evaluate importance of system resources (e.g., files, registry, and processes) and malware detection. While, Egele \emph{et al.} \cite{Egele2014} presented a dynamic blanket execution function that employs high-level API-relevant semantic features. Enck \emph{et al.} \cite{Enck2014} presented \emph{TaintDroid} for dynamic taint examination to trace leakage of sensitive data (e.g., microphone, GPS and camera) in third-party apps. Ye \emph{et al.} \cite{Ye2017} presented a deep learning strategy comprised of AutoEncoder, multilayer restricted Boltzmann machines and associative memory. The framework detects malware in embedded systems via Windows API calls extricated from portable executable files. 

Though dynamic analysis techniques are independent of malware source-code and can detect unknown and zero-day malware instances, they require more resources (e.g., memory, CPU time and disk space) and have high computational cost and false positive rates.

\subsubsection{Hybrid Analysis}
It is also called gray-box approach. Neither static- nor dynamic-analysis methods are unable to provide perfect anti-malware solutions. Thus, hybrid-analysis approaches, which combine benefits of both static and dynamic analyses, is more desirable. For instance, Santos \emph{et al.} \cite{Santos2012} designed a hybrid method that integrates static (i.e., opcodes frequency) and dynamic (i.e., executable’s execution trace data) features with multitude of classifiers. Authors in \cite{Tong2017} proposed a hybrid technique that collects system calling runtime data and then utilizes a static scheme for mobile malware detection. While, Dali \emph{et al.} \cite{Dali2017} developed a method that uses FlowDroid static analysis tool and sensitive sources data flows with deep learning-based classifier.

\subsection{Feature selection}
The performance of malware detection depends on choice of feature representation and length. The feature selection/dimensionality reduction is conducted to attain a set of more discriminative features for enhanced performance. Various anti-malwares have been presented using filter, wrapper and embedding based feature selection algorithms such as distributed-, hierarchical-, correlation-, low-rank matrix approximation-, forward-, backward-, local sensitive hashing-, max relevance, adaptive feature scaling-, spectral graph theory-, F1-score, F2-score, mean decrease impurity-, document frequency-, information gain-, information gain ratio-, principal component analysis- and latent dirichlet allocation \cite{Souri2018}.

\subsection{Classification/Clustering}
To identify if a given sample is malicious or/and to determine malware family, various binary and multiclass classification techniques such as Multilayer Perceptron, Support Vector Machines, Naïve Bayes, Decision Tree, Rule-based, Random Forests, Multiple Kernel Learning, $K$-Nearest Neighbors, Logistic Regression, Ensemble, Multi-Objective Evolutionary by Genetic Algorithm, Deep Belief Networks have been employed \cite{Ucci2017}.

Hierarchical-, $K$-means-, meanShift-, $K$-medoid partitional-, density-based spatial-, prototype-, self-organizing maps-, single-linkage- and locality sensitive hashing-based clustering techniques have been utilized to categorize malware samples exhibiting identical behaviors into different groups or to generate signatures for detection \cite{Souri2018}.

\subsection{Evaluation Metrics}
Performance of malware detection methods is generally evaluated by False Positive Rate = FP/(FP + TN), True Positive Rate = TP/(TP + FN), specificity = TN/(TN + FP), precision = TP/(TP + FP), accuracy =  (TP + TN)/(TP + TN + FP + FN),  where TP, FP, TN and FN are true positives, false positives, true negatives and false negatives, respectively. Malware samples are commonly considered as positive instances. Moreover, Matthews correlation coefficient, F-score, Kappa statistic, confusion matrix, receiver operating characteristic and under the curve measures have been used. While, for clustering-based algorithms Macro-F1 and Micro-F1 metrics, respectively, for accentuating the performance on rare and common categories \cite{Souri2018,Ucci2017}.

\section{Research challenges and opportunities}
\label{sec:researchchallengesandopportunities}
The ever-growing demand of minimized failure rates of anti-malware solutions have opened up exigent research opportunities and challenges to be resolved yet.

\subsection{Issues in existing anti-malware methods}
Malwares are still exponentially evolving in sophistication, and more difficult plights lie ahead. Most prior static and dynamic or hybrid methods do not work for novel/unknown/zero-day signatures and require virtual environment plus are time consuming, respectively. Nonetheless, virtual environments are becoming less effective as malware writers are usually one step ahead by implementing high-level new techniques to conceal malicious features. Though efforts are afoot to design multi-level and parallel processing system, existing anti-malware methods/tools all in all are not adequate or potent for higher levels of concealments. Current anti-malware systems also face challenges like scalability, lack of truly real-world representative datasets, irreproducibility of published results, low generalization and detection disagreement among them for the same samples. There is a need of improved and comprehensive malware countermeasures, which could be developed by utilizing recent advanced-machine/deep learning, -data mining and -adaptive schemes. Also, approaches embodying anomaly analysis with behavioral data should be designed to investigate what the malware is doing rather than how it is doing. This may result in minimized error and false alarm rates.

\subsection{Advanced machine learning (AML) techniques for anti-malware}
Quintessential anti-malwares often depend on non-linear adversary explicit models and expert domain knowledge, thereby making them prone to overfitting and lower overall reliability. Conversely, AML techniques attempt to imitate attackers with various content, contexts, etc. rather than explicit models/systems/attacks. Few preliminary studies on shallow AML usage for anti-malware has been conducted, but still a lot of efforts to be done regarding AML anti-malware. For improved accuracy, flexibility and scalability on wide range and unknown samples, AML paradigms like open set recognition, more complex and residual deep learning, dictionary learning and data mining should be explored for feature segmentation/representation learning/selection/classification and determining temporal relationships within and between malware sections.

\subsection{Mobile device malwares}
Smart-devices connected to internet is growing exponentially, so as malwares (especially via third party apps) against them. Insubstantial studies have been conducted on mobile device malwares. Moreover, most existing anti-malware techniques are not real-time and unsuited for mobile devices because of high computational cost and/or features complexity used for analysis. Thus, real-time lightweight mobile anti-malwares via Bayesian classification is an interesting research direction to be explored. Multiple information from in-built sensors (e.g., accelerometer) may enhance mobile anti-malware performance. Mobile hardware malware detection and removal is another issue that needs serious exploration. Sooner mobile anti-malware-inspired techniques will substantially impact smart-devices design. Anyway, smart-device malwares should be tackled both by preventive and effective countermeasures. App developers should assure that their apps are abiding security and privacy policies. App stores administrators should vet and remove dubious apps. Users should use superior anti-malwares and install trusted apps. On the whole, wearable and mobile devices malware and anti-malware are a new research field in cybersecurity with pressing problems worth researching like malware affecting device's master boot record or stealthily exploiting device to mine cryptocurrency, and how a malware performing well on benchmark data will be better under real-world environments. 

\subsection{Large-scale benchmark databases}
Advancement in malware research deeply depends on the public availability of comprehensive benchmark datasets incorporating accurate labels and contexts. Most existing databases suffers from limitations like small size, missing information/features, imbalanced-classes, and not publicly available. Lack of adequate large-scale public datasets has stymied research on malware. Benchmark public datasets will assist to compare independent anti-malware schemes, determine inter and intra relationships between security infringement phenomena and unify malware findings to draw determined conclusions with reference to statistical significance. Nevertheless, collecting large-scale heterogenous annotated databases is challenging and time- and resource-consuming due to malware attributes, forms and behaviors diversity. Crowdsourcing may help accumulating different annotated large-scale databsets. 

\subsection{Graph-based malware analysis}
Malwares with concealments are dominant nowadays and effectual in evading conventional anti-malwares that largely disregard learning and identifying the underlying relationships between samples and variants, and contextual information. Graph-based relationship representations and features (e.g., data- and control-flow graphs, call graphs, data-, program-, and control-dependency graphs) offer interesting possibility even when malware code is altered as it helps in tracking malware genealogy in different settings. Devising graph-based anti-malwares yet have issues from data heterogeneity, noisy and incomplete labels, and computational cost during real-time detection. Up to some extend such challenges may be addressed in decentralized fashion. Furthermore, use of multiple directed and undirected graphs, multi-view spectral clustering, heterogeneous networks, multiple graph kernel learning, dynamic graph mining and deep graph convolution kernels to capture contextual and structural information could be fruitful area of research.

\subsection{Bio-inspired anti-malware}
Several limitations of traditional anti-malwares could be suppressed by bio-inspired (e.g., biological immune system, biological evolution, genetic algorithms and swarm intelligence) techniques. Comparatively these techniques are lightweight, highly scalable and less resource-constrained. Adaptive bio-inspired techniques that is used both for intelligent concealment-invariant feature extraction and classification can dramatically enhance accuracy in the wild. Bio-inspired methods that define particular objective functions to discriminate a system under attack from a malfunctioning or failing may also help strengthening the security. Combination of bio-inspired algorithms with deep neural networks is one of the most promising direction, however has been explored less in anti-malwares.

\subsection{Defense-in-depth anti-malware}
Anti-malware strategy that is composed of multiple defense levels/lines rather than single is called defense-in-depth. Such strong defensive mechanism is expected to be more robust as it doesn’t depend on one defense technique and if one is breached the others aren’t. Each machine/cyber-system architecture can be divided in various levels of depth, e.g., in a power grid system, the meters, communication frameworks, and smaller components, respectively, could be envisaged as lowest, intermediate and highest level. Another solution is active or adaptive malware defense. Active defense has received little attention due to inherent complexity, where developer anticipates attack scenarios at different level and accordingly devises malware countermeasures. In adaptive defense, the system is persistently updated by retraining/appending novel features or dynamically adjusted corresponding to reshaping environments. Adaptive defenses would require fast, automated and computationally effective and could use unsupervised learning and concept drift.

\subsection{Internet of things (IoT) attacks}
IoT are progressively being used in different domains ranging from smart-cities to smart- and military-grids. Despite finest security endeavors, IoT devices/systems can also be compromised by innovative cyber-attacks. Security of IoT technology is more crucial as it is always connected to a network. IoT cyber-security is relatively new research realm and quite challenging owing to heterogeneous networks with multisource data and several categories of nodes.  To this end, different routes (e.g., predictive and blockchain) could be effective. Predictive security is attaining cyber resiliency by devising models that predict future attacks and prevent in advance. As there is a strong correlation between security infractions and human blunders, predictive models should consider computer networking, social sciences, descriptive theory, uncertain behavior theory and psychology from attackers, users and administrators' perspectives at different granularity levels. Blockchain can be utilized for self-healing of compromised devices/systems. Models could be devised that exploit e.g., redundancy to heal corrupted codes/software by good codes replacements, since in blockchain one can trace and roll back the firmware versions. However, such models should also be capable to handle resource, energy and communication constraints, which may be achieved by lightweight machine/transfer/reinforcement learning based access control protocols. 

\subsection{Deception and moving target anti-malware techniques}
Deception techniques (e.g., honeypot) are being used to detect and prevent malwares, which lures adversaries to strike in order to mislead them with false information. There are two kinds of honeypots, i.e., client and server. Honeypot helps to reduce false positives and prevent DDoS attacks. Complex attacks/tools (e.g., polymorphic malware) is increasing to identify honeypots or to alter their behaviors to deceive honeypots themselves. Also, honeypot can be exploited by attackers to undermine other sensitive parts of frameworks. More complicated honeypot and honey nets (i.e., bunch of honeypots) schemes (e.g., shadow honeypots) should be devised as compromised honeypot will put security of whole organization in danger.

Moving target techniques (aka dynamic platform methods-DPMs) dynamically randomizes system components to suppress successful attacks’ likelihood and shorten attack lifetime. Though adversary must undermine all platforms not one to evade DPMs, DPMs require complicated application state synchronization among varying platforms, and expand the system’s attack surface. Much less efforts have been dedicated to developing well-articulated attack models and how to upgrade deception elements and strategy to confront dynamic changes in attack behaviors. Future research should concentrate on devising unorthodox methodologies, performing real-world analyses to compute and compare effectiveness of deception and DPMs techniques, and studying if DPMs conflict or can co-exist with other anti-malwares.

\subsection{Decentralized anti-malware}
Data sharing and trust management hinder current anti-malwares advancement, which can be resolved by decentralized malware detectors using blockchain technology. But it has received little attention till now. For intersection of anti-malware and blockchain technology, future directions will include exploring overhead traffic handling, quality and sparse malware signatures, building accurate dynamic normal nature of traffics, reducing massive false alerts, energy and cost, blockchain latency, case-by-case scenario investigation, and more proof-of-concept implementations.

\subsection{Botnet countermeasures}
Thwarting botnets has become key area. Several botnet detection and defense architectures have been proposed. Various issues surround botnet countermeasure study, e.g., difficulties in testing devised botnet defenses in real scenarios/data. Besides, lack of widely acknowledged benchmark or standard methodology to quantitative evaluate or compare bot defenses presumably due to privacy and data sharing concerns. Botnets, including IoT bot and socialbot, will continue to rise until effective means both technical and non-technical are taken. Technical factors include passive internet service providers and unassertive software. Non-technical factors include establishing distributed global environment, local and multinational legal issues and poor user awareness.

\subsection{Privacy preservation}
Malwares that steal sensitive information has received much attention. However, preserving user privacy in malware analysis (especially at the cloud or third party server) and malware data sharing is yet an open and seldom touched concern. Establishing privacy and regaining trust in commercial anti-malwares would become difficult if user's privacy/data is compromised once. Majority of prior anti-malwares overlook the privacy and security of user, data and network. Thus, reasonably little has been worked on privacy protection frameworks to respect public and law opinions. Privacy preservation mechanisms that do not influence the detection performance is practically worthy of contemplation. Formulating lightweight detection and privacy protection systems usable on mobile devices to balance security, efficacy, privacy and power consumption demands special considerations. More innovative privacy preservation approaches (e.g., allowing user to stabilize privacy, convenience and security levels) in malware analysis has been highlighted by many experts as an essential future research to be carried out.

\subsection{Big data malware analysis}
The demand for big data malware analysis frameworks is steadily expanding. Practitioners are working to resolve big data malware challenges such as volume (e.g., collecting, cleaning and compressing data/labels), velocity (e.g., real-time online training, learning, processing or streaming big data), variety (e.g., heterogeneous multi-view data learning/embedding), veracity (e.g., leaning with contradicting and unreliable data), and value (explainable ML based malware analysis). Another promising future research direction is devising large-scale feature selection techniques, which are less-dependent on feature engineering, via distributed feature selection, low-rank matrix approximation, adaptive feature scaling, spectral graph theory, and fuzzy and neuro-fuzzy clustering. Rigorous efforts need to be made to investigate use of synchronous parallel processing (e.g., Spark) and to develop body of knowledge on pros and cons of big data anti-malware tools to assist practitioners.

\subsection{Malware analysis visualization systems}
Existing methods to analyze malwares are time-consuming for malware analysts. Highly interactive visual analysis would aid researchers and practitioners to forensically investigate, summarize, classify and compare malwares more easily. Most prior techniques are very limited with regard to interactivity, mapping temporal dimensions, scalability and representation space (e.g., they are superficially 2D rather than 3D). The field of developing malware visualization systems covering consequential rang of malware types and environments is vital and emerging. Encyclopedic visualization systems will lead analysts/researchers to ascertain novel research domains in the years to come.

\subsection{Multimodal anti-malwares}
Multimodal anti-malwares, which consolidate evidences from different kinds of features/sources (e.g., string, permission, elements converted to image matrices) can overcome numerous constraints in frameworks that consider only one/fewer features. Multimodal frameworks are more flexible and can significantly enhance the accuracy of unimodal ones in the wild. Multimodal may include multiple sensors, algorithms and instances, and information can be fused at feature, score or decision level. There is ample room to develop novel fusion architectures. Moreover, multimodal frameworks are expected to be intrinsically more robust to concealments, but no study investigated how robust are they to concealments.

\subsection{Clustering for malware analysis}
Previous works have shown that clustering could be a useful tool to effectively classify unknown malwares for improved generalization, to underline unseen family’s behaviors for thorough analysis that may help more robust anti-malware schemes, and to label huge volumes of malwares in fast and automatic fashion that has become major challenge. Future goal should be further improving accuracy of clustering-based malware analysis using cluster quality measurements, contextual/metadata information, and boosted genetic algorithms, etc. Attentions should also be given to rectify security issues, e.g., poisoning and obfuscation attacks against targeted clusters. 

\subsection{Hardware-based Solutions}
Hardware-based detectors are recently getting momentum against proliferation of malware. Such detection mechanisms utilize low-level architectural features, which are obtained by redesigning the micro-architecture of computer processors, e.g., CPUs with special registers providing hardware and software anomaly events. Nevertheless, research in this domain and trustworthy systems (i.e., inherently secure and reliable against human errors and hostile parties) is yet in its initial genesis and has to go a long way. Furthermore, there is dearth of studies on efficacy of anti-malwares combining hardware- and software-based techniques that have exceptional potential to uncover extra elaborate malwares. Likewise, smart devices’ sensors (e.g., GPS and ambient light sensors) data could also be used as additional feature vector to profile malware. 

\subsection{Malware adversarial learning}
Machine-learning (ML) recently has been used to achieve effective malware defenses, however they are not designed for situations where an adversary is actively trying to impact outcomes. Specially, deep learning-based countermeasures lack robustness against adversarial examples (i.e., sample crafted from genuine samples with careful minor perturbations). Attackers can also inject poisoning samples in (online/adaptive) training database with the aim to remarkably decrease ML-malware countermeasure’s accuracy at testing phase. A comprehensive analysis for each malware considering attacker’s capability and what features to what extend should be modified to avoid detection has not been done yet. It is still difficult task to design ML anti-malwares that are robust in adversarial setting. Researchers should explore malware adversarial ML in identifying probable countermeasures' vulnerabilities both at train and test stages, devising homologous assails and their impacts and developing techniques to enhance robustness of ML-based anti-malwares. 

\subsection{Performance evaluation framework}
Malware analysis accuracy/performance, which is used to evaluate, compare, or configure anti-malwares, in general lacks standardization. A unified and comprehensive evaluation framework should be developed to rank present and future methods, that incorporates static and dynamic techniques, adversary's goal, knowledge and capability, attack strategies at train and test phase, evaluation metrics (i.e., security- and privacy-relevant error rates as most current methods do not cover all aspects), and common parlance to elucidate anti-malware performances. Any such framework with common criteria and open online platform for evaluating resilient, malware sophistication, decision making, policies, experimental setups, big databases, and open source codes will surely help both in reporting baseline performances without giving a false sense of progress and encouraging reproducible research on scalability and challenges in real-world scenarios. 

\subsection{Malware education}
Most malwares succeed contemplating humans as weakest link. Additionally, there is growing demand for cybersecurity workforces, therefore it is imperative to educate people about malware safety. In academic institutions, malware analysis and related courses should be taught both at undergraduate and graduate levels. Nonetheless, relatively very limited colleges/universities offer malware courses, which maybe because of the shortage of agreement on fundamental topics among institutions, book and training providers, and ethical sensitivity of educating/creating white-hats. Moreover, most academic courses being offered are practitioner-oriented but not science-/research-oriented and heavily rely on text books that are not current. Some training camps/workshops are being held by companies/organizations also for general public, but they are exceptionally expensive. More on-line free-to-access training courses will surely diminish malware damages. 

\subsection{Interdisciplinary research}
To advance state-of-the-art malware analysis, the research and industrial communities need to support and promote interdisciplinary fundamental science research and development (including contributions from machine learning, human psychology, computer engineering, etc.) to accomplish dependable, natural, and generalized anti-malware techniques.

\section{Conclusion}
Malwares, including in mobile and smart devices, have become more sophisticated and greater in frequency during recent years. Although there exist many defense tools and mechanisms, malware detection and analysis are still challenging tasks, since malware developers continuously conceal the information in attacks or evolve cyber-attacks to circumvent newer security techniques, plus some prior methods face low generalization to unknown malwares and scalability issues. It is hoped that this academic and perspective article will stimulate focused interdisciplinary research and development in anti-malware towards aggrandizing its full potential in different cyberspace applications.

\end{document}